\documentclass[letterpaper, 10 pt, conference]{ieeeconf}  % Comment this line out
\IEEEoverridecommandlockouts                              % This command is only
\usepackage{graphics} % for pdf, bitmapped graphics files
\usepackage{epsfig} % for postscript graphics files
\usepackage{amsfonts,amsmath}
\usepackage{amssymb}

\newtheorem{thm}{Theorem}%
\newtheorem{lem}{Lemma}
\newtheorem{rem}{Remark}
\newtheorem{cor}{Corollary}

\newcommand{\Id}{{1\!\!\!1}}

\newcommand{\EE}[1]{{\text{\large$\mathbb E$}}\left(#1\right)}
\newcommand{\PP}[1]{{\text{\large$\mathbb P$}}\left(#1\right)}

\newcommand{\MM}{{\mathcal M}}
\newcommand{\WWW}{{\mathcal W}}

\newcommand{\bra}[1]{\left<#1\right|}
\newcommand{\ket}[1]{\left|#1\right>}
\newcommand{\tr}[1]{\text{Tr}\left(#1\right)}
\newcommand{\bket}[1]{\left<#1\right>}

\newcommand{\dotex}{{\frac{d}{dt}}}
\newcommand{\half}{\text{\scriptsize $\frac{1}{2}$}}

\newcommand{\rhoe}{\rho^{\text{\tiny est}}}

\newcommand{\nmax}{n^{\text{\tiny max}}}

\title{Feedback generation of quantum Fock states by discrete QND measures
    \thanks{This work was   supported in part by the "Agence Nationale de la Recherche" (ANR),
    Projet Blanc  CQUID number 06-3-13957.}}

\author{ Mazyar Mirrahimi, Igor Dotsenko and Pierre Rouchon% <-this % stops a space
\thanks{M. Mirrahimi is with INRIA Rocquencourt,
        Domaine de Voluceau, B.P. 105, 78153 Le Chesnay cedex, France,
        {\tt\small mazyar.mirrahimi@inria.fr}}
\thanks{I. Dotsenko is with Laboratoire Kastler Brossel, \'Ecole Normale Sup\'erieure, CNRS,
        Universit\'{e} P. et M. Curie, 24 rue Lhomond, F-75231 Paris Cedex 05, France
        {\tt\small igor.dotsenko@lkb.ens.fr}}%
\thanks{P. Rouchon is with Mines ParisTech,  Centre Automatique et Syst\`emes, Math\'{e}matiques et Syst\`{e}mes,
        60 Bd Saint Michel, 75272 Paris cedex 06, France,
        {\tt\small pierre.rouchon@mines-paristech.fr}}
}

\begin{document}

\maketitle \thispagestyle{empty} \pagestyle{empty}

%%%%%%%%%%%%%%%%%%%%%%%%%%%%%%%%%%%%%%%%%%%%%%%%%%%%%%%%%%%%%%%%%%%%%%%%%%%%%%%%
\begin{abstract}
A feedback scheme for preparation  of photon number states  in a microwave cavity is proposed. Quantum Non Demolition (QND) measurement of the cavity field provides information on its actual state. The control consists in injecting into the cavity mode a  microwave pulse adjusted to increase the population of the desired target photon number. In  the ideal case (perfect cavity and measures), we present  the  feedback scheme  and  its  detailed convergence proof through stochastic Lyapunov techniques based on super-martingales and  other probabilistic arguments. Quantum Monte-Carlo simulations performed with experimental parameters illustrate   convergence and robustness of such feedback scheme.
\end{abstract}

%%%%%%%%%%%%%%%%%%%%%%%%%%%%%%%%%%%%%%%%%%%%%%%%%%%%%%%%%%%%%%%%%%%%%%%%%%%%%%%%
\section{Introduction}\label{sec:intro}

In~\cite{deleglise-et-al:nature08,guerlin-et-al:nature07,gleyzes-et-al:nature07} QND  measures are exploited to detect and/or produce highly non-classical states of light trapped  in a super-conducting cavity
(see~\cite[chapter 5]{haroche-raimond:book06} for a  description of such QED  systems and~\cite{brune-et-al:PhRevA92} for detailed physical models with QND measures of light using atoms). For such experimental setups, we detail and analyze  here a  feedback scheme that stabilize the cavity field towards any photon-number  states (Fock states). Such states are  strongly  non-classical since  their photon numbers are   perfectly defined. The  control corresponds to a  coherent light-pulse injected inside the cavity between atom passages. The overall structure of the proposed feedback scheme is inspired of~\cite{geremia:PRL06} using a  quantum adaptation of the observer/controller structure  widely used for classical  systems (see, e.g.,~\cite[chapter 4]{kailath-book}). The observer part of the proposed feedback scheme consists in a discrete-time quantum filter, based on the observed detector clicks,  to estimate the  quantum-state of the cavity field.  This estimated state is then used in a  state-feedback based on Lyapunov design, the controller part. In theorems~\ref{thm:main} and~\ref{thm:initial} we prove the convergence almost surely of the closed-loop system towards the goal Fock-state in absence of modeling imperfections  and measurement errors. Simulations  illustrate this results  and show that performance  of the closed-loop system are not dramatically changed by false detections for  $10\%$ of the detector clicks. In~\cite{dotsenko-et-al:PRA09} similar  feedback schemes are also addressed with  modified quantum filters in order to take into account additional physical effects and  experimental imperfections. \cite{dotsenko-et-al:PRA09} focuses on physics  and includes extensive closed-loop simulations whereas here we are interested by  mathematical aspects and  convergence proofs.

In section~\ref{sec:ideal}, we describe very briefly  the  physical system  and its quantum Monte-Carlo  model. In section~\ref{sec:feedback} the feedback is designed using Lyapunov techniques. Its convergence is proved in theorem~\ref{thm:main}. Section~\ref{sec:filtering} introduces a quantum filter to estimate the cavity state necessary for the feedback: convergence of the closed-loop system (quantum filter and feedback based on the estimate cavity state) is proved in theorem~\ref{thm:initial} assuming perfect model and detection.This section ends with Theorem~\ref{thm:contract} proving a contraction property of the quantum filter dynamics.  Section~\ref{sec:simul} is devoted to closed-loop simulations where measurement imperfections are introduced for testing robustness.

The authors thank  Michel Brune, Serge Haroche and Jean-Pierre Raimond for useful discussions and advices.

\section{The physical system and its jump dynamics}\label{sec:ideal}
\begin{figure}[htb]
 \centerline{\includegraphics[width=0.45\textwidth]{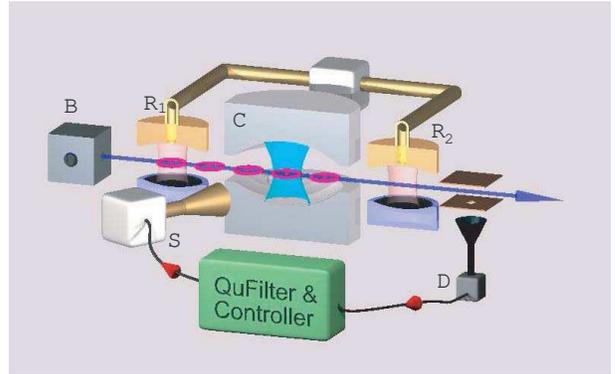}}
  \caption{The microwave cavity QED setup with its  feedback scheme (in green).}\label{fig:ExpScheme}
\end{figure}
As illustrated by figure~\ref{fig:ExpScheme}, the system consists in $C$ a high-Q microwave cavity, in $B$ a box producing Rydberg atoms, in $R_1$ and $R_2$ two low-Q Ramsey
cavities, in $D$ an atom detector and in $S$ a microwave source.
%We use   to model the measurement and control process associated with the experimental setup %(see~\cite[chapter 4]{haroche-raimond:book06}).
%Note that, although quantum trajectories represent a useful simulation method
%for the quantum master equation, they are interesting in their own right, since they
%model the measurement process itself and the resulting conditioned dynamics. In fact,
%some of the original work that motivated quantum trajectories~\cite{cohenT-dalibard:86,zoller-et-al-87} was to understand
%experiments on quantum jumps (for atoms with a vee configuration)~\cite{nagourney-et-al-86,sauter-et-al-86}.
The dynamics model is discrete in time and  relies on quantum Monte-Carlo trajectories (see~\cite[chapter 4]{haroche-raimond:book06}). It  takes  into account the back-action of the measure. It is adapted from~\cite{guerlin-et-al:nature07} where we have just added the control effect.

 Each time-step indexed by the integer $k$ corresponds to  atom number $k$ coming from $B$, submitted then to a first Ramsey $\pi/2$-pulse in $R_1$,  crossing the cavity $C$  and being entangled with it, submitted to a second $\pi/2$-pulse in $R_2$  and finally being measured in $D$. The state of the cavity is described  by the density operator $\rho_{k}$. Here the passage from the time step $k$ to $k+1$ corresponds to the projective measurement of the cavity state, by detecting the state of the Rydberg atom number $k$ after leaving $R_2$. During this same  step, an appropriate coherent pulse (the control)  is injected into $C$. Denoting, as usual, by $a$ the photon annihilation operator and by $N=a^\dag a$
the photon number operator, the density matrix $\rho_{k+1}$ is related to $\rho_k$ through the following jump-relationships:
$
    \rho_{k+1} =\frac{ D(\alpha_{k}) M_k  \rho_k  M_k^\dag D(-\alpha_{k})}{\tr{M_k \rho_k M_k^\dag}}
$
where
\begin{itemize}
\item the measurement operator $M_k=M_g$ (resp. $M_k=M_e$), when the atom $k$ is detected in the state $\ket{g}$ (resp. $\ket{e}$)  with
{\small \begin{align}\label{MgMe:eq}
M_g = \cos\left(\frac{\phi_R+\phi}{2} + N \phi \right),~
M_e = \sin\left(\frac{\phi_R+\phi}{2} + N \phi \right).
\end{align}}
Such measurement process corresponds to an off-resonant interaction between atom $k$ and cavity where $\phi_R$ is the direction of the second Ramsey $\pi/2$-pulse ($R_2$ in figure~\ref{fig:ExpScheme}) and $\phi$ is the  de-phasing-angle  per photon.

\item The probability  $P_{g,k}$ (resp. $P_{e,k}$) of detecting the atom $k$ in $\ket{g}$ (resp. $\ket{e}$) is given by $\tr{M_g\rho_k M_g}$ (resp.  $\tr{M_e\rho_k M_e}$.

\item $D(\alpha_k)$ is the displacement operator describing the coherent pulse injection, $D(\alpha_k)=\exp(\alpha_k (a^\dag - a))$,   and the scalar control $\alpha_k$ is  a real parameter that can be  adjusted at each time step~$k$.

\end{itemize}
The time evolution of the step $k$ to $k+1$, in fact, consists of two types of evolutions: a projective measurement and a coherent injection. For simplicity sakes, we will use the notation of $\rho_{k+\half}$, to illustrate this intermediate step. Therefore,
{\small\begin{align}\label{idealdyn:eq}
    \rho_{k+\half} =\frac{  M_k  \rho_k  M_k^\dag }{\tr{M_k \rho_k M_k^\dag}}, \quad
        \rho_{k+1}=D(\alpha_{k}) \rho_{k+\half} D(-\alpha_{k})
\end{align}}

In the sequel, we consider the finite dimensional  approximation defined by a maximum of photon number, $\nmax$. In the truncated Fock basis $\left(\ket{n}\right)_{0\leq n \leq\nmax}$, $N$ corresponds to the diagonal matrix $\left(\text{diag}(n)\right)_{0\leq n \leq\nmax}$, $\rho$ is a $(\nmax+1)\times(\nmax+1)$ symmetric  positive matrix with unit trace, and the annihilation operator $a$ is an upper-triagular matrix with  $\left(\sqrt{n}\right)_{1\leq n \leq\nmax}$ as  upper diagonal,  the remaining elements being $0$. We restrict to real quantities since the phase of any Fock state is arbitrary. We  set it  here to $0$.

\section{Feedback scheme and convergence proof}\label{sec:feedback}

We aim to stabilize the Fock state with $\bar n$ photons characterized by the density operator $\bar\rho=\ket{\bar n}\bra{\bar n}$. To this end we choose the coherent feedback $\alpha_k$ such that the value of the Lyapunov function $V(\rho)=1-\tr{\rho\bar\rho}$ decreases when passing from $\rho_{k+\half}$ to $\rho_{k+1}$. Note that, for $\alpha$ small enough, the Baker-Campbell-Hausdorff formula yields the following approximation
{\small \begin{align}\label{BKH:eq}
D(\alpha) \rho D(-\alpha)
\approx \rho -\alpha [\rho,a^\dag-a] + \frac{\alpha^2}{2}[[\rho,a^\dag-a],a^\dag-a]
\end{align}}
up to third order terms.
Therefore, for $\alpha_k$ small enough, we have
{\small \begin{multline*}
\tr{D(\alpha_k) \rho_{k+\half} D(-\alpha_k) \bar \rho}= \\\tr{\rho_{k+\half} \bar\rho}
- \alpha_k \tr{[\rho_{k+\half},a^\dag-a]\bar \rho} \\
+ \frac{\alpha_k^2}{2}\tr{[[\rho_{k+\half},a^\dag-a],a^\dag-a]\bar\rho}.
\end{multline*}}
Thus the feedback
\begin{equation}\label{eq:feed1}
\alpha_k= c_1 \tr{[\bar\rho ,a^\dag-a]\rho_{k+\half}}
\end{equation}
with a gain $c_1 >0$ small enough ensures that
{\small \begin{equation}\label{eq:lyap}
\tr{\bar\rho\rho_{k+1} } - \tr{\bar\rho\rho_{k+\half} } \geq  \frac{c_1}{2} \Big|\tr{[\bar\rho ,a^\dag-a]\rho_{k+\half}}\Big|^2,
\end{equation}}
since $\tr{[\rho_{k+\half},a^\dag-a]\bar \rho}=-\tr{[\bar \rho, a^\dag-a]\rho_{k+\half}}$.
Furthermore, the conditional expectation of $\tr{\bar\rho \rho_{k+\half}}$ knowing $\rho_k$ is given by
{\small \begin{multline*}
\EE{\tr{\bar\rho\rho_{k+\half}}~|~\rho_k} = P_{g,k} \tr{\frac{\bar \rho M_g\rho_k M_g^\dag}{P_{g,k}}}\\
+ P_{e,k} \tr{\frac{\bar \rho M_e\rho_k M_e^\dag}{P_{e,k}}} = \tr{\bar \rho \rho_k}
\end{multline*}}
since $[\bar \rho, M_g]=[\bar \rho, M_e]=0$ and $M_g^\dag M_g+M_e^\dag M_e=\Id$.  Thus
$$
\EE{\tr{\bar\rho\rho_{k+1}}~|~\rho_k} \geq \EE{\tr{\bar\rho\rho_{k+\half}}~|~\rho_k}=\tr{\bar\rho \rho_k}
$$
and consequently, the expectation value of $V(\rho_k)$ decreases at each sampling time:
\begin{equation}\label{eq:martingale}
\EE{V(\rho_{k+1})} \leq \EE{V(\rho_k)}.
\end{equation}
Considering the Markov process $\rho_k$, we have therefore shown that $V(\rho_k)$ is a super-martingale bounded from below by 0.
When $V$ reaches its minimum $0$, the Markov process $\rho_k$ has converged to $\bar\rho$. However, one can
easily see that this super-martingale has also the possibility to converge towards other attractors, for instance other Fock states which are all the stationary points of the closed-loop Markov process but with $V(\rho)=1$ instead of $0$. Following~\cite{mirrahimi-handel:siam07}, we suggest the following modification of the feedback scheme: \small
\begin{equation}\label{eq:feedmain}
    \alpha_k=\left\{
    \begin{array}{ll}
    c_1 \tr{[\bar\rho ,a^\dag-a]\rho_{k+\half}}
        & \mbox{ if } V(\rho_{k})\le 1-\varepsilon\\
    \underset{\alpha\in[-\bar\alpha,\bar\alpha]}{\text{argmax}}~\tr{\bar\rho D(\alpha)\rho_{k+\half}D(-\alpha)} & \mbox{ if } V(\rho_{k})>1-\varepsilon \\
    \end{array}
    \right.
\end{equation}\normalsize
with $c_1,\varepsilon,\bar\alpha>0$  constants.

\begin{thm}\label{thm:main}
Consider~\eqref{idealdyn:eq} and assume that for all $n\in\{0,\ldots,\nmax\}$ we have
$\frac{\phi_R+\phi}{2} + n \phi\neq 0 \mod(\pi/2)$ and that
{\small $$
\# \left\{
\cos^2\left(\frac{\phi_R+\phi}{2} + n \phi \right)~|~ n\in\{0,\ldots,\nmax\}
\right\} = \nmax+1.
$$}
Take the switching feedback scheme~\eqref{eq:feedmain} with $\bar\alpha >0$.  For  small enough $c_1 >0$ and  $\varepsilon>0$, the trajectories of~\eqref{idealdyn:eq} converge almost surely towards the target Fock state $\bar\rho$.
\end{thm}
\begin{rem}\label{rem:support}
The second part of the feedback~\eqref{eq:feedmain}, dealing with states near the bad attractors, is not explicit and may seem hard to compute. Note that, this form has been particularly chosen to simplify the proof of the Theorem~\ref{thm:main} and in practice, one can take it to be any constant control field exciting the system around these bad attractors and ensuring a fast return to the inner set.
 %All the simulations of section~\ref{sec:robustness} are performed with such constant control laws % (cf. feedback law~\eqref{eq:feedexp}).
\end{rem}
\begin{rem}\label{rem:gain}
The controller gain $c_1$ can be tuned in order to maximize at each sampling time $k$, $\tr{D(\alpha_k) \rho_{k+\half} D(-\alpha_k) \bar \rho}$ for $ \rho_{k+\half}$ near $\bar\rho$. Up to third order term in $\rho_{k+\half}-\bar\rho$, \eqref{BKH:eq} yields to\small
\begin{multline*}
\tr{D(\alpha_k) \rho_{k+\half} D(-\alpha_k) \bar \rho}
= \tr{\bar\rho \rho_{k+\half}}
+\\
\left(\tr{[\bar\rho,a^\dag-a]\rho_{k+\half}}\right)^2 \left( c_1 - \frac{c_1^2}{2} \tr{[\bar\rho,a^\dag-a][\bar\rho,a^\dag-a]}\right).
\end{multline*}\normalsize
Thus $ c_1 = 1/\tr{[\bar\rho,a^\dag-a][\bar\rho,a^\dag-a]} \approx 1/(4\bar n+2)$ for $\nmax \gg \bar n$
implies a maximum decrease at the sampling time, up to third-order terms in $\rho_k-\bar\rho$.
\end{rem}
In order to prove the Theorem~\ref{thm:main}, we need some classical tools from stochastic processes namely the Doob's inequality and the Kushner's asymptotic invariance Theorem~\cite{kushner-71}. These results are been recalled in the Appendix.

\textit{Proof of Theorem~\ref{thm:main}.}
It is divided in 3 steps: in a first step, we show that for small enough $\varepsilon$ and by applying the second part of the feedback scheme, the trajectories starting within the set $\{\rho~|~ V(\rho)>1-\varepsilon\}$ reach in one step the set $\{\rho~|~ V(\rho)\leq 1-2\varepsilon\}$ and this with probability 1; next, we show that  trajectories starting within the set $\{\rho~|~ V(\rho)\leq 1-2\varepsilon\}$, will never hit the set $\{\rho~|~ V(\rho)>1-\varepsilon\}$ with a uniformly non-zero probability $p>0$; finally, we will show that, the trajectories of the quantum filter converge towards $\bar\rho$ for almost all trajectories that never hit the set $\{\rho~|~ V(\rho)>1-\varepsilon\}$. This is then an immediate conclusion of the Markov property that the trajectories of the quantum filter with the feedback scheme~\eqref{eq:feedmain} will converge almost surely towards $\bar\rho$.

\text{Step 1:} We start by considering the process starting on the level set $\{\rho~|~V(\rho)=1\}$. We have the following lemma:
\begin{lem}\label{lem:step1}
Consider $\rho$ a well-defined density matrix such that $\tr{\rho\bar\rho}=0$. We have
{\small \begin{equation*}
\underset{s\in\{g,e\}}{\min}\underset{\alpha\in[-\bar\alpha,\bar\alpha]}{\max}~\frac{\tr{\bar\rho D(\alpha)M_s\rho M^{\dag}_s D(-\alpha)}}{\tr{M_s\rho M^{\dag}_s}}>0.
\end{equation*}}
We denote any  argument of the above min-max problem by $\bar\alpha(\rho)\in[-\bar\alpha,\bar\alpha]$.
\end{lem}
\textit{Proof of Lemma~\ref{lem:step1}:}
Define
$
\rho_s=\frac{M_s\rho M^{\dag}_s}{\tr{M_s\rho M^{\dag}_s}},\quad s\in\{g,e\}.
$
The matrices $M_g$ and $M_e$ being diagonal, we trivially have
$\tr{\rho_s\bar\rho}=0$. Let us fix $s$ and assume that for all $\alpha\in[-\bar\alpha,\bar\alpha]$,
\begin{equation}\label{eq:absurd}
\tr{\bar\rho D(\alpha)\rho_s D(-\alpha)}=0.
\end{equation}
Decomposing $\rho_s$ as a sum of projectors we have
$
\rho_s=\sum_{k=1}^m \lambda_{k,s} \ket{\psi_{k,s}}\bra{\psi_{k,s}},
$
where $\lambda_{k,s}$ are strictly positive eigenvalues and $\psi_{k,s}$ are the associated normalized eigenstates of $\rho_s$ ($m=1$ corresponds to the case where $\rho_s$ is a projector). The equation~\eqref{eq:absurd}, clearly, implies
{\small\begin{equation}\label{eq:absurd2}
\bket{\psi_{k,s}~|~D(-\alpha)\bar n}=0,\quad \forall k\in\{1,\cdots,m\},\forall \alpha\in[-\bar\alpha,\bar\alpha].
\end{equation}}
Fixing one $k\in\{1,\cdots,m\}$ and taking $\psi=\psi_{k,s}$,  noting that $D(-\alpha)=\exp(-\alpha(a^\dag-a))$ and deriving $j$ times versus $\alpha$ around $0$  we get
\begin{equation}\label{eq:absurd3}
\bket{\psi~|~(a^\dag-a)^j\bar n}=0,\quad \forall j=0,\ldots,\nmax.
\end{equation}
But the family $\left((a^\dag-a)^j\bar n\right)_{0\le j \le \nmax}$ is full rank.
This is a direct consequence of~\cite[Theorem~4]{schirmer-et-al:PRA}.  It is proved there that the truncated harmonic oscillator $ \dotex \ket{\phi}_t = -(\imath N + v(t) (a^\dag - a)) \ket{\phi}_t,$   is completely controllable with the single scalar control $v(t)$. If the rank $r$ of  $\left((a^\dag-a)^p\ket{\bar n}\right)_{0\le p \le \nmax}$ is  strictly less that  $\nmax+1$, then according to Cayley-Hamilton Theorem the rank of the infinite family $\left((a^\dag-a)^p\ket{\bar n}\right)_{p\geq 0}$  is also $r$. Take $\ket{\xi}$, a state  orthogonal to this family.  For any  control $v(t)$,  the state $\ket{\phi}_t$ starting from $\ket{\bar n}$ remains orthogonal to $\ket{\xi}$. Thus it will be impossible to find a control $v(t)$ steering $\ket{\phi}_t$ from $\ket{\bar n}$ to $\ket{\xi}$.

Since the rank of $\left((a^\dag-a)^p\ket{\bar n}\right)_{0\le p \le \nmax}$ is maximum, \eqref{eq:absurd3} implies $\ket{\psi_k}=0$ and leads to a contradiction. $\square$

Applying the compactness of the space of density matrices, we directly have the following corollary:
\begin{cor}\label{cor:step1}
There exists an $\epsilon>0$ such that
{\small \begin{equation}\label{eq:cor}
\underset{\rho\in\{\tr{\rho\bar\rho}<\epsilon\}}{\inf}~\frac{\tr{\bar\rho D(\bar\alpha(\rho))M_s\rho M^{\dag}_s D(-\bar\alpha(\rho))}}{\tr{M_s\rho M^{\dag}_s}}>2\epsilon
\end{equation}}
for $s=g,e$ and where $\bar\alpha(\rho)$ is defined in Lemma~\ref{lem:step1}.
\end{cor}
\textit{Proof of Corollary~\ref{cor:step1}:} We take
{\small $$
\delta=\underset{\rho\in\{\tr{\rho\bar\rho}=0\}}{\inf}\underset{s\in\{g,e\}}{\min}~\frac{\tr{\bar\rho D(\bar\alpha(\rho))M_s\rho M^{\dag}_s D(-\bar\alpha(\rho))}}{\tr{M_s\rho M^{\dag}_s}}.
$$}
By Lemma~\ref{lem:step1} and the compactness of the set $\{\rho~|~\tr{\rho\bar\rho}=0\}$, we know that
$\delta>0$. By continuity of $\tr{\rho\bar\rho}$ with respect to $\rho$ and by compactness of the space of density matrices, there exists $\gamma>0$ such that
{\small $$
\underset{\rho\in\{\tr{\rho\bar\rho}<\gamma\}}{\inf}\underset{s\in\{g,e\}}{\min}~\frac{\tr{\bar\rho D(\bar\alpha(\rho))M_s\rho M^{\dag}_s D(-\bar\alpha(\rho))}}{\tr{M_s\rho M^{\dag}_s}}>\frac{\delta}{2}.
$$}
Therefore, by taking $\epsilon=\min(\gamma,\delta/4)$, clearly,~\eqref{eq:cor} holds true. $\square$

Through this corollary, we have shown that whenever the Markov process hits the set $\{\tr{\rho\bar\rho}<\epsilon\}$, it is immediately rebounded to the set
$\{\tr{\rho\bar\rho}>2\epsilon\}$  and this with probability~1.

\textit{Step 2:} Let us  assume that the process starts within the set $\{\tr{\rho\bar\rho}>2\epsilon\}$.
\begin{lem}\label{lem:step2}
Initializing the Markov process within the set $\{\rho~|~V(\rho)\leq 1-2\epsilon\}$, $\rho_k$ will never hit the set $\{\rho~|~V(\rho)> 1-\epsilon\}$ with a probability
$
p>\frac{\epsilon}{1-\epsilon}>0.
$
\end{lem}
\textit{Proof of Lemma~\ref{lem:step2}:} By~\eqref{eq:martingale}, the process $V(\rho_k)$ is, clearly, a supermartingale. One only needs to use the Doobs inequality (cf. Appendix, Theorem~\ref{thm:doob}) and we have
{\small $$
P(\underset{0\leq k<\infty}{\sup} V(\rho_k)> 1-\epsilon)< \frac{V(\rho_0)}{1-\epsilon}\leq \frac{1-2\epsilon}{1-\epsilon},
$$}
and thus $p> 1-(1-2\epsilon)/(1-\epsilon)=\epsilon/(1-\epsilon)$. $\square$

We have shown that starting within the inner set $\{\tr{\rho\bar\rho}\geq 2\epsilon\}$ there is a uniform non-zero probability of $\epsilon/(1-\epsilon)$ for the process, to never hit the outer set $\{\tr{\rho\bar\rho}< \epsilon\}$.

\textit{Step 3:}
\begin{lem}\label{lem:step3-1}
The sample paths $\rho_k$ remaining into the set $\{\tr{\rho\bar\rho}>\epsilon\}$ converge in probability to $\bar\rho$ as $k\rightarrow\infty$.
\end{lem}
\textit{Proof of Lemma~\ref{lem:step3-1}:} Consider the function
$
\WWW(\rho)=1-\tr{\rho\bar\rho}^2.
$
For $s=g,e$, set
$
\rho_s=\frac{M_s\rho M_s^\dag}{\tr{M_s\rho M_s^\dag}}$.
We have
{\small \begin{align}\label{eq:W1}
\WWW(\rho_g)&=1-\frac{\tr{\rho M_g^\dag\bar\rho M_g}^2}{\tr{M_g\rho M_g^\dag}^2},\notag\\
&=1-\frac{\Big|\cos\left(\frac{\phi_R+\phi}{2}+\bar n\phi\right)\Big|^4}{\tr{M_g\rho M_g^\dag}^2}\tr{\rho\bar\rho}^2,
\end{align} }
and similarly
{\small \begin{equation}\label{eq:W2}
\WWW(\rho_e)=1-\frac{\Big|\sin\left(\frac{\phi_R+\phi}{2}+\bar n\phi\right)\Big|^4}{\tr{M_e\rho M_e^\dag}^2}\tr{\rho\bar\rho}^2.
\end{equation}}
Furthermore, whenever $\alpha$ is given by the first  part of the feedback law, we have
{\small \begin{equation}\label{eq:W3}
\WWW(D(\alpha)\rho D(-\alpha))-\WWW(\rho)\le  -2\epsilon c_1 \Big|\tr{[\bar\rho ,a^\dag-a]\rho}\Big|^2,
\end{equation}}
where we have applied~\eqref{eq:lyap} together with the fact that
$$
|\tr{D(\alpha)\rho D(-\alpha)\bar\rho}|+|\tr{\rho\bar\rho}|\ge 2\epsilon
$$
since $\rho$ is inside the set $\{\tr{\rho\bar\rho}>\epsilon\}$.
Applying~\eqref{idealdyn:eq},~\eqref{eq:W1},~\eqref{eq:W2} and~\eqref{eq:W3} for the paths never leaving the set $\{\tr{\rho\bar\rho}>\epsilon\}$, we have
\footnotesize
\begin{multline*}
\EE{\WWW(\rho_{k+1})~|~\rho_k}-\WWW(\rho_k) \le -2\epsilon c_1 \Big|\tr{[\bar\rho ,a^\dag-a]\rho_{k+\half}}\Big|^2 \\
-\text{\footnotesize $\left(\frac{\Big|\cos\left(\frac{\phi_R+\phi}{2}+\bar n\phi\right)|^4}{\tr{M_g\rho_k M_g^\dag}}+\frac{\Big|\sin\left(\frac{\phi_R+\phi}{2}+\bar n\phi\right)|^4}{\tr{M_e\rho_k M_e^\dag}}-1\right)$}\tr{\rho_k\bar\rho}^2.
\end{multline*}\normalsize
Noting that$\tr{M_g\rho_k M_g^\dag}\ge 0$, $\tr{M_e\rho_k M_e^\dag}\ge 0$,
$\tr{M_g\rho_k M_g^\dag}+\tr{M_e\rho_k M_e^\dag}=1$
and by Cauchy-Schwartz inequality, we have
{\small \begin{multline*}
\frac{\Big|\cos\left(\frac{\phi_R+\phi}{2}+\bar n\phi\right)\Big|^4}{\tr{M_g\rho_k M_g^\dag}}+\frac{\Big|\sin\left(\frac{\phi_R+\phi}{2}+\bar n\phi\right)\Big|^4}{\tr{M_e\rho_k M_e^\dag}}=\\
\left(\frac{\Big|\cos\left(\frac{\phi_R+\phi}{2}+\bar n\phi\right)\Big|^4}{\tr{M_g\rho_k M_g^\dag}}+\frac{\Big|\sin\left(\frac{\phi_R+\phi}{2}+\bar n\phi\right)\Big|^4}{\tr{M_e\rho_k M_e^\dag}}\right)\\(\tr{M_g\rho_k M_g^\dag}+\tr{M_e\rho_k M_e^\dag})\geq \\
\left(\cos^2\left(\frac{\phi_R+\phi}{2}+\bar n\phi\right)+\sin^2\left(\frac{\phi_R+\phi}{2}+\bar n\phi\right)\right)^2=1,
\end{multline*}}
with equality if and only if
$
\tr{M_g\rho_k M_g^\dag}=\cos^2\left(\frac{\phi_R+\phi}{2}+\bar n\phi\right).
$
We apply, now, the Kushner's invariance Theorem (cf. Appendix, Theorem~\ref{thm:kushner}) to the Markov process $\rho_k$ with the Lyapunov function $\WWW(\rho_k)$. The process $\rho_k$ converges in probability to the largest invariant set included in
\begin{multline*}
\Big\{\rho~|~ \tr{M_g\rho M_g^\dag}=\cos^2\left(\frac{\phi_R+\phi}{2}+\bar n\phi\right)\Big\}\\
\bigcap \left\{\rho~|~\tr{[\bar\rho ,a^\dag-a]M_s\rho M_s^\dag}=0,~s=g,e\right\}.
\end{multline*}
In particular, by invariance, $\rho$ belonging to this limit set implies
$
\tr{M_g\rho M_g^\dag}=\frac{\tr{M_g M_s\rho M_s^\dag M_g^\dag}}{\tr{M_s\rho M_s^\dag}}$ for $ s=g,e$.
Taking $s=g$, and noting that $M_g=M_g^\dag$, this leads to
$
\tr{M_g^4\rho}=\tr{M_g^2\rho}^2.
$
However, by Cauchy-Schwartz inequality, and applying the fact that $\rho$ is a positive matrix, we have
$
\tr{M_g^4\rho}=\tr{M_g^4\rho}\tr{\rho}\geq \tr{M_g^2\rho}^2,
$
with equality if and only if $M_g^4\rho$ and $\rho$ are co-linear. Since $M_g^4$ has a non degenerate spectrum, $\rho$ is  necessarily a projector over one of the eigen-state of $M_g^4$, i.e.,   a  Fock state $\ket{n}$, for some $n\in\{0,\ldots,\nmax\}$.
Finally, as we have restricted ourselves to the paths never leaving the set $\{\rho~|~\tr{\rho\bar\rho}>\epsilon\}$, the only possibility for the invariant set is the isolated point $\{\bar\rho\}$. $\square$
\begin{lem}\label{lem:step3-2}
$\rho_k$ converges to $\bar\rho$ for almost all paths remaining in the set $\{\tr{\rho\bar\rho}>\epsilon\}$.
\end{lem}
\textit{Proof of Lemma~\ref{lem:step3-2}:} Define the event $P_{>\epsilon}=\{\omega\in\Omega~|~\rho_k \text{ never leaves the set } \{\tr{\rho\bar\rho}>\epsilon\}\}$. Through Lemma~\ref{lem:step3-1}, we have shown that
$
\lim_{k \rightarrow\infty}\PP{\|\rho_k-\bar\rho\|>\delta~|~P_{>\epsilon}}=0,$ $\forall\delta>0.
$
By continuity of $V(\rho)=1-\tr{\rho\bar\rho}$, this also implies that
$
\lim_{k \rightarrow\infty}\PP{V(\rho_k)>\delta~|~P_{>\epsilon}}=0,$ $\forall\delta>0.
$
As $V(\rho)\leq 1$, we have
\begin{multline*}
\EE{V(\rho_k)~|~P_{>\epsilon}}\leq \PP{V(\rho_k)> \delta~|~P_{>\epsilon}}\\
+\delta(1-\PP{V(\rho_k)> \delta~|~P_{>\epsilon}}).
\end{multline*}
Thus
$
\limsup_{k\rightarrow\infty} \EE{V(\rho_k)~|~P_{>\epsilon}}\leq \delta,$ $\forall\delta>0,
$
and so
$
\lim_{k\rightarrow\infty} \EE{V(\rho_k)~|~P_{>\epsilon}}=0.
$
By Theorem~\ref{thm:doob}, we know that $V(\rho_k)$ converges almost surely and therefore, as $V$ is bounded,
by dominated convergence, we obtain
$
\EE{\lim_{k\rightarrow\infty} V(\rho_k)~|~P_{>\epsilon}}=0.\square
$

Now, we have all the elements to finish the proof of the Theorem~\ref{thm:main}. From Steps 1 and 2 and the Markov property, one deduces that for almost
all paths $\rho_k$, there exists a $\bar K$ such that $\rho_k$ for $k\geq \bar K$ never leaves the set $\{\tr{\rho\bar\rho}>\epsilon\}$. This together with the step 3 finishes the proof of the Theorem.$\square$

\section{Quantum filtering for state estimation} \label{sec:filtering}
The feedback law~\eqref{eq:feedmain} requires the knowledge of $\rho_{k+\half}$.
When the  measurement process is fully efficient and the jump model~\eqref{idealdyn:eq} admits no error, it actually represents a natural choice for  quantum filter to estimate the value of $\rho$ by  $\rhoe$  satisfying
\begin{align}\label{idealfilter:eq}
    \rhoe_{k+1}&=D(\alpha_{k}) \rhoe_{k+\half} D(-\alpha_{k})\notag\\
    \rhoe_{k+\half} &=\frac{  M_{s_k}  \rhoe_k  M_{s_k}^\dag }{\tr{M_{s_k} \rhoe_k M_{s_k}^\dag}}.
\end{align}
where $s_k=g$ or $e$, depending on  measure outcome $k$ and on the control $\alpha_k$.

Before passing to the parametric robustness of the feedback scheme, let us discuss the robustness with respect to the choice of the initial state for the filter equation when we replace $\rho_{k+\half}$ by $\rhoe_{k+\half}$ in the feedback~\eqref{eq:feedmain}.   Note that, Theorem~\ref{thm:main} shows that whenever the filter equation is initialized at the same state as the one which the physical system is prepared initially, the feedback law ensures the stabilization of the target state. The next theorem shows that as soon as the quantum filter is initialized at any arbitrary fully mixed initial state (not necessarily the same as the initial state of the physical system~\eqref{idealdyn:eq}) and whenever the  feedback scheme~\eqref{eq:feedmain} is applied on the system, the state of the physical system will converge almost surely to the desired Fock state.
\begin{thm}\label{thm:initial}
Assume that the quantum filter~\eqref{idealfilter:eq} is initialized at a full-rank matrix $\rhoe_0$ and that the feedback scheme~\eqref{eq:feedmain} is applied to the physical system. The trajectories of the system~\eqref{idealdyn:eq}, will then converge almost surely to the target Fock state $\bar\rho$.
\end{thm}
\textit{Proof of Theorem~\ref{thm:initial}:} The initial state $\rhoe_0$ being full-rank, there exists a $\gamma>0$ such that
$
\rhoe_0=\gamma\rho_0+(1-\gamma)\rho_0^c,
$
where $\rho_0$ is the initial state of~\eqref{idealdyn:eq} at which the physical system is initially prepared and $\rho_0^c$ is a well-defined density matrix. Indeed, $\rhoe$ being positive and full-rank, for a small enough $\gamma$, $(\rhoe_0-\gamma\rho_0)/(1-\gamma)$ remains non-negative,  Hermitian and of unit trace.

Assume that, we prepare the initial state of another identical physical system as follows: we generate a random number $r$ in the interval $(0,1)$; if $r<\gamma$ we prepare the system in the state $\rho_0$ and otherwise we prepare it at $\rho_0^c$. Applying our quantum filter~\eqref{idealfilter:eq} (initialized at $\rhoe_0$) and the associated feedback scheme, almost all trajectories of this physical system converge to the Fock state $\bar\rho$. In particular, almost all trajectories that were initialized at the state $\rho_0$ converge to $\bar\rho$. This finishes the proof of the theorem. $\square$

The quantum filter~\eqref{idealfilter:eq} admits  also some contraction properties confirming  its robustness to experimental errors as shown by simulations of figures~\ref{fig:TrajErr} and~\ref{fig:Mean} where detection errors are introduced.  We just provide here a first interesting inequality that will be used  in future developments.
\begin{thm}\label{thm:contract}
Consider the process~\eqref{idealdyn:eq} and the associated filter~\eqref{idealfilter:eq} for any arbitrary control input $(\alpha_k)_{k=1}^\infty$. We have
$
\EE{\tr{\rho_k\rhoe_k}}\leq \EE{\tr{\rho_{k+1}\rhoe_{k+1}}},$ $\forall k.
$
\end{thm}
\textit{Proof}
Before anything, note that the coherent part of the evolution leaves the value of $\tr{\rho_k\rhoe_k}$ unchanged: {\small \begin{multline*}
\tr{\rho_{k+1}\rhoe_{k+1}}=
\tr{D(\alpha_k)\rho_{k+\half}\rhoe_{k+\half}D(-\alpha_k)}\\=
\tr{\rho_{k+\half}\rhoe_{k+\half}}.
\end{multline*}}
Concerning the projective part of the dynamics, we have
{\small
\begin{multline}\label{eq:C1}
\EE{\tr{\rho_{k+\half}\rhoe_{k+\half}}~|~\rho_k,\rhoe_k}=\\
\sum_{s=g,e} \frac{\tr{M_s\rho_k M_s^\dag M_s \rhoe_k M_s^\dag}}{\tr{M_s\rhoe_k M_s^\dag}}.
\end{multline}}
Applying a Cauchy-Schwarz inequality as well as the identity $M_g^\dag M_g+M_e^\dag M_e=\Id$, we have
{\small
\begin{multline}\label{eq:C2}
\sum_{s=g,e} \frac{\tr{M_s^\dag M_s \rho_k M_s^\dag M_s \rhoe_k}}{\tr{M_s\rhoe_k M_s^\dag}}=\\
\sum_{s=g,e} \tr{M_s\rhoe_k M_s^\dag} \sum_{s=g,e} \frac{\tr{M_s^\dag M_s\rho_k M_s^\dag M_s \rhoe_k}}{\tr{M_s\rhoe_k M_s^\dag}}\geq\\
\left(\sum_{s=g,e} \sqrt{\tr{M_s^\dag M_s\rho_k M_s^\dag M_s \rhoe_k}}\right)^2
\end{multline}}
Applying~\eqref{eq:C1} and~\eqref{eq:C2}, we only need to show that
{\small
\begin{multline}\label{eq:C3}
\left(\sum_{s=g,e} \sqrt{\tr{M_s^\dag M_s\rho_k M_s^\dag M_s \rhoe_k}}\right)^2 \geq \tr{\rho_k\rhoe_k}.
\end{multline}}
Noting, once again, that $M_g^\dag M_g+M_e^\dag M_e=\Id$, we can write:
\begin{equation}\label{eq:C4}
\tr{\rho_k\rhoe_k}=\sum_{s=g,e}\sum_{r=g,e} \tr{M_{s}^\dag M_{s} \rho_k M_{r}^\dag M_{r} \rhoe_k},
\end{equation}
and therefore~\eqref{eq:C3} is equivalent to
{\small
\begin{multline}\label{eq:C5}
\sum_{s=g,e} \sqrt{\tr{M_{s}\rho_k M_s^\dag M_s \rhoe_k M_s^\dag}} \sum_{r=g,e} \sqrt{\tr{M_{r}\rho_k M_r^\dag M_r \rhoe_k M_r^\dag}}\\
\geq \sum_{s=g,e}~\sum_{r=g,e}\tr{M_s^\dag M_s \rho_k M_r^\dag M_r \rhoe_k}.
\end{multline}}
Note that as $\rho_k$ and $\rhoe_k$ are positive Hermitian matrices, their square roots, $\sqrt{\rho_k}$ and $\sqrt{\rhoe_k}$, are well-defined. Once again by Cauchy-Schwarz inequality, we have
{\small
\begin{align*}
&\tr{M_s^\dag M_s \rho_k M_r^\dag M_r \rhoe_k}=
\tr{\sqrt{\rhoe_k} M_s^\dag M_s \sqrt{\rho_k}  \sqrt{\rho_k}  M_r^\dag M_r \sqrt{\rhoe_k}}\\
&\leq\sqrt{\tr{\sqrt{\rhoe_k} M_s^\dag M_s \sqrt{\rho_k}  \sqrt{\rho_k}  M_s^\dag M_s \sqrt{\rhoe_k}}} \\
&\qquad\qquad\qquad\qquad \qquad\sqrt{\tr{\sqrt{\rhoe_k} M_r^\dag M_r \sqrt{\rho_k}  \sqrt{\rho_k}  M_r^\dag M_r \sqrt{\rhoe_k}}}\\
&=\sqrt{\tr{M_s\rho_k M_s^\dag M_s\rhoe_k M_s^\dag}}\sqrt{\tr{M_r\rho_k M_r^\dag M_r\rhoe_k M_r^\dag}}.
\end{align*}}
Summing over $s,r\in\{g,e\}$, we obtain the inequality~\eqref{eq:C5} and therefore we finish the proof of the Theorem~\ref{thm:contract}.  $\square$

\section{Monte-Carlo simulations}\label{sec:simul}

\begin{figure}
  \centerline{ \includegraphics[width=0.5\textwidth]{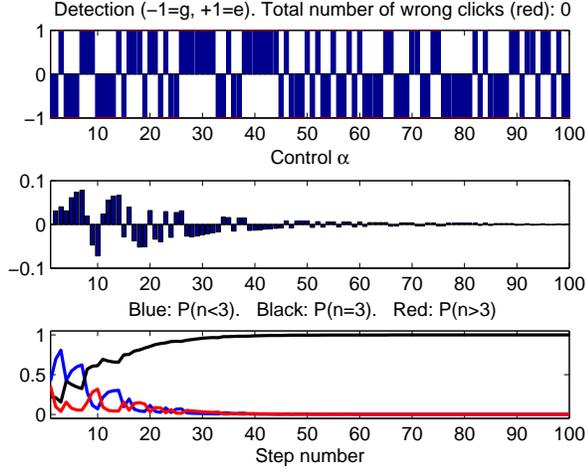}}
  \caption{A single closed-loop  quantum trajectory  in the ideal case ($\bar n=3$). }\label{fig:TrajIdeal}
\end{figure}

Figure~\eqref{fig:TrajIdeal} corresponds to a closed-loop simulation with  a goal Fock state
$\bar n=3$  and  a Hilbert space  limited to $\nmax = 15$ photons. $\rho_0$ and $\rhoe_0$ are initialized at the same state, the coherent state $\exp(\sqrt{\bar n}(a^\dag-a))\ket{0}$ of mean photon number $\bar n$. The number of iteration steps is fixed to $100$. The dephasing per photon is $\phi=\frac{3}{10}$.  The Ramsey phase $\phi_R$ is fixed to the mid-fringe setting, i.e. $\frac{\phi_R+\phi}{2}+\bar n\phi=\frac{\pi}{4}$.  The feedback parameter (\eqref{eq:feedmain} with $\rhoe_{k+\half}$ instead of $\rho_{k+\half}$) are as follows:  $c_1=\frac{1}{4\bar n +1}$, $\epsilon=\frac{1}{10}$ and $\bar\alpha=\frac{1}{10}$.

\begin{figure}
\centerline{\includegraphics[width=0.5\textwidth]{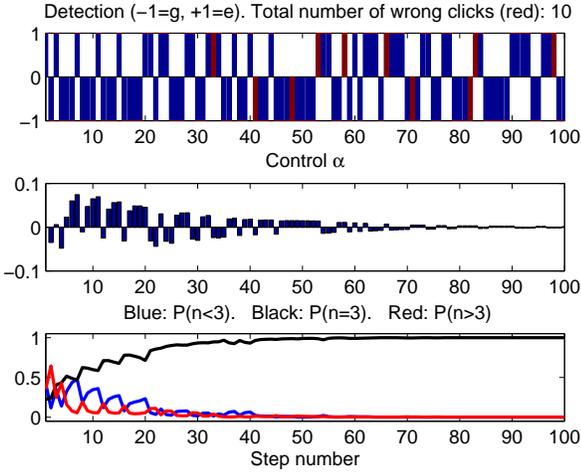}}
  \caption{A single closed-loop  quantum trajectory  with a false detection probability of $1/10$. }\label{fig:TrajErr}
\end{figure}

\begin{figure}
   \centerline{\includegraphics[width=0.25\textwidth]{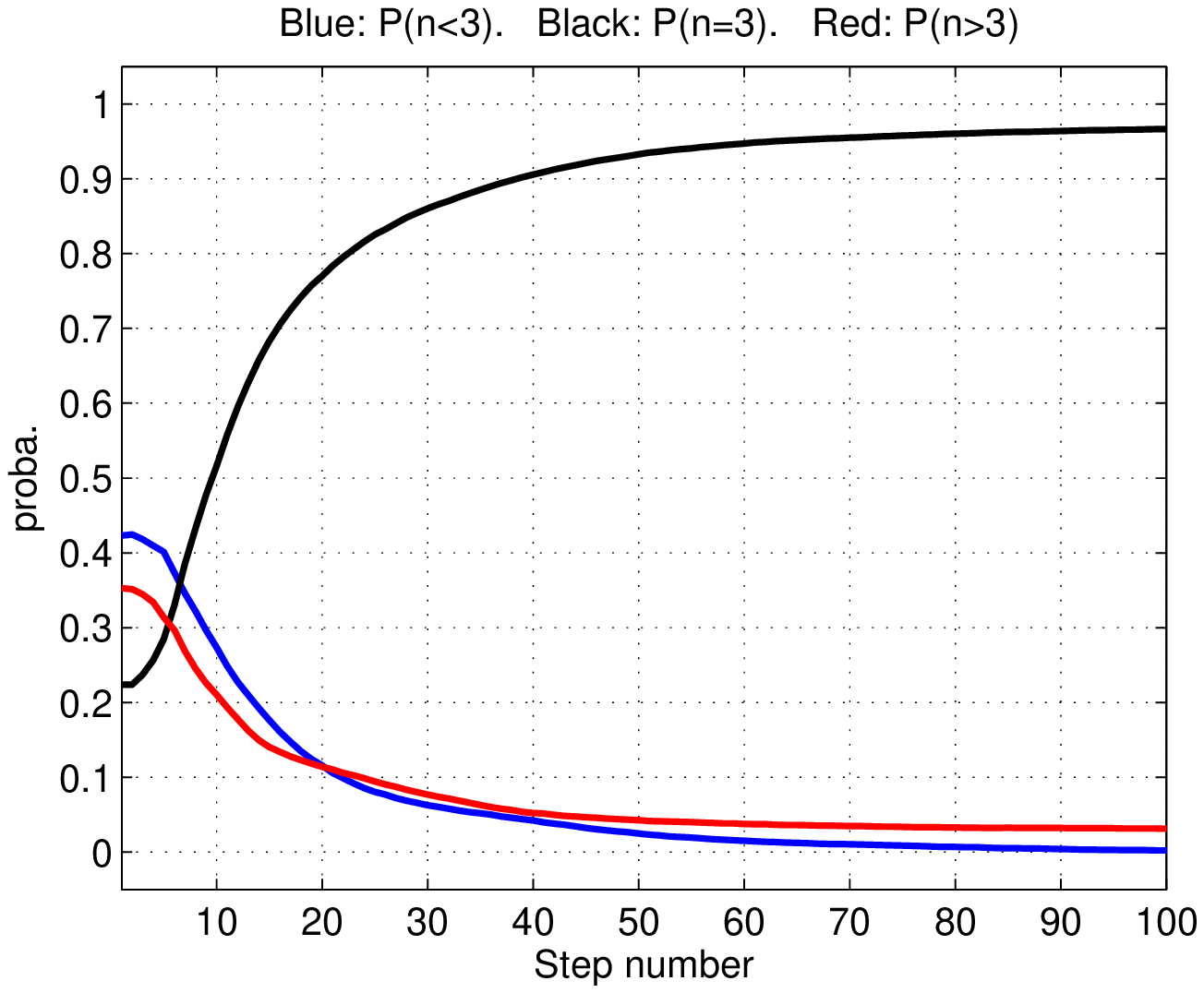}
        \includegraphics[width=0.25\textwidth]{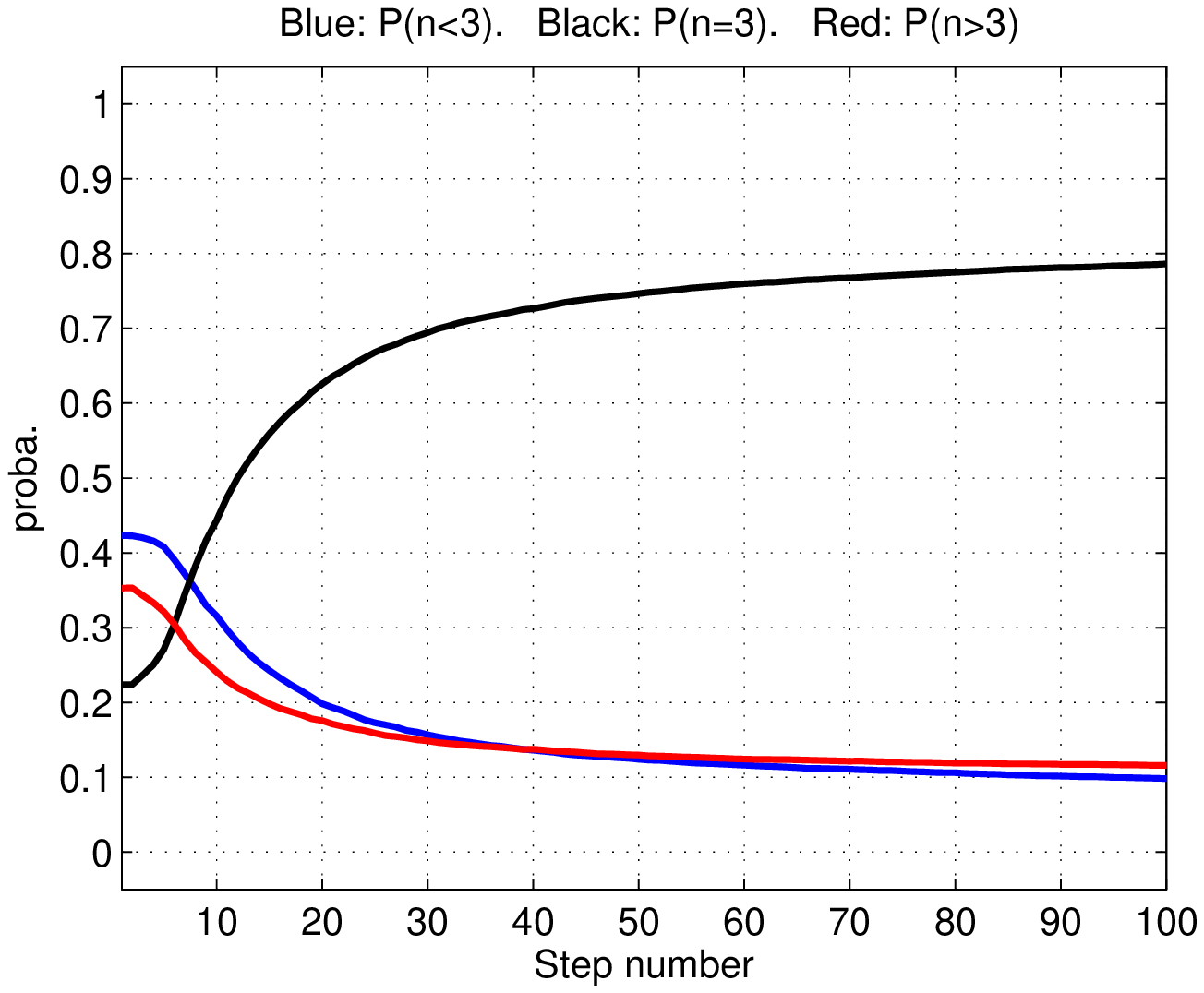}}
  \caption{Averages of $10^4$ closed-loop  quantum trajectories  similar to the one of figure~\ref{fig:TrajIdeal} (left, $\eta_f=0$) and~\ref{fig:TrajErr} (right, $\eta_f=\frac{1}{10}$). }\label{fig:Mean}
\end{figure}

Any real experimental setup includes imperfection and error. To test the robustness of the feedback scheme,  a  false detection probability $\eta_{f}=\frac{1}{10}$  is introduced. In case of false detection at step $k$, the atom  is detected in $g$ (resp. $e$) whereas it collapses effectively  in $e$ (resp. $g$). This means that in~\eqref{idealfilter:eq}, $s_k=g$ (resp. $s_k=e$), whereas in~\eqref{idealdyn:eq}, it is the converse $M_k=M_e$ (resp. $M_k=M_g$).
Simulations of figure~\ref{fig:TrajErr} differ  from those of  figure~\ref{fig:TrajIdeal} by only $\eta_f=\frac{1}{10}$: we observe for this sample trajectory a longer convergence time.
A much more significative impact of $\eta_f$ is given by ensemble average.
Figure~\ref{fig:Mean} presents   ensemble averages corresponding to  the third sub-plot of figures~\ref{fig:TrajIdeal} and~\ref{fig:TrajErr}.  For $\eta_f=0$ (left plot), we observe an average fidelity $\tr{\rho_k\bar\rho}$  converging to $100\%$: it exceeds $90\%$  after $k=40$ steps. For $\eta_f=1/10$, the asymptotic fidelity remains under $80\%$  and reaches  $70\%$ after $30$ iteration. The performance are reduced but not changed dramatically. The proposed feedback scheme appears to be  robust  to such experimental errors.

\section{Conclusion}\label{sec:conclusion}

In~\cite{dotsenko-et-al:PRA09} more realistic simulations are reported. They  include nonlinear shift per photon ($N\phi$ replaced by a non linear function $\Phi(N)$  in~\eqref{MgMe:eq}) and additional  experimental errors  such as detector efficiency and delays. These  simulations confirm the robustness of the feedback scheme, robustness that needs to be understood in a more theoretical way. In particular, it seems that the quantum filter~\eqref{idealfilter:eq}  forgets its initial condition $\rhoe_0$ almost surely  and thus admits some strong contraction properties as indicated by Theorem~\ref{thm:contract}.

With the truncation to $\nmax$ photons, convergence is proved only in the finite dimensional case. But feedback~\eqref{eq:feedmain} and quantum filter~\eqref{idealfilter:eq} are still valid for  $\nmax=+\infty$. We conjecture that  Theorems~\ref{thm:main} and~\ref{thm:initial} remain valid in this case.

In the experimental results  reported in~\cite{deleglise-et-al:nature08,guerlin-et-al:nature07,gleyzes-et-al:nature07} the  time-interval corresponding to  a sampling step is around $100\mu s$. Thus it is   possible  to implement, on a digital computer and  in real-time, the  Lyapunov feedback-law~\eqref{eq:feedmain} where $\rho$ is given by the quantum filter~\eqref{idealfilter:eq}.

\section{Appendix: stability theory for stochastic processes}
We recall here Doob's inequality and  Kushner's invariance theorem. For detailed discussions and proofs we refer to~\cite{kushner-71} (Sections 8.4 and 8.5).
\begin{thm}[Doob's Inequality]\label{thm:doob}
Let $\{X_n\}$ be a Markov chain on state space $S$. Suppose that there is a non-negative function $V(x)$ satisfying
$
\EE{V(X_1)~|~X_0=x}-V(x)=-k(x),
$
where $k(x)\geq 0$ on the set $\{s:V(x)<\lambda\}\equiv Q_\lambda$. Then
$
\PP{\underset{\infty>n\geq 0}{\sup} V(X_n)\geq \lambda~|~X_0=x}\leq \frac{V(x)}{\lambda}.
$
Furthermore, there is some random $v\geq 0$, so that for paths never leaving $Q_\lambda$, $V(X_n)\rightarrow v\geq 0$ almost surely.
\end{thm}
For the statement of the second Theorem, we need to use the language of probability measures rather than the random process.
Therefore, we deal with the space $\MM$ of probability measures on the state space $S$. Let $\mu_0=\varphi$ be the initial probability distribution (everywhere through this paper we have dealt with the case where $\mu_0$ is a dirac on a state $\rho_0$ of the state space of density matrices). Then, the probability distribution of $X_n$, given initial distribution $\varphi$, is to be denoted by $\mu_n(\varphi)$. Note that for $m\geq 0$, the Markov property implies:
$
\mu_{n+m}(\varphi)=\mu_n(\mu_m(\varphi)).
$
\begin{thm}[Kushner's invariance Theorem]\label{thm:kushner}
Consider the same assumptions as that of the Theorem~\ref{thm:doob}. Let $\mu_0=\varphi$ be concentrated on a state $x_0\in Q_\lambda$  ($Q_\lambda$ being defined as in Theorem~\ref{thm:doob}), i.e. $\varphi(x_0)=1$.  Assume that $0\leq  k(X_n)\rightarrow 0$ in $Q_\lambda$ implies that $X_n\rightarrow \{x~|~k(x)=0\}\cap Q_\lambda\equiv K_\lambda$. Under the conditions of Theorem~\ref{thm:doob}, for trajectories never leaving $Q_\lambda$, $X_n$ converges to $K_\lambda$ almost surely. Also, the associated conditioned probability measures $\tilde\mu_n$ tend to the largest invariant set of measures $M$ whose support set is in $K_\lambda$. Finally, for the trajectories never leaving $Q_\lambda$, $X_n$ converges, in probability, to the support set of $M$.
\end{thm}

\end{document}